\documentstyle[mprocl]{article}
\begin{document}

\title{A PHENOMENOLOGICAL APPROACH TO QUINTESSENCE\\
IN SCALAR-TENSOR GRAVITY}

\author{G. ESPOSITO-FARESE}

\address{Centre de Physique Th\'eorique, CNRS Luminy,\\
Case 907, F 13288 Marseille cedex 9, France\\
and D\'epartement d'Astrophysique Relativiste et de Cosmologie,\\
Observatoire de Paris-Meudon, F 92195 Meudon cedex, France\\
E-mail: gef@cpt.univ-mrs.fr}

\maketitle\abstracts{This talk is based on my work in collaboration
with B.~Boisseau, D.~Polarski, and A.A.~Starobinsky. The most natural
and best-motivated alternatives to general relativity are the
so-called ``scalar-tensor'' theories, in which the gravitational
interaction is mediated not only by a (spin-2) graviton, but also by
a (spin-0) scalar field. We study quintessence in this general
framework, and show that the microscopic Lagrangian of the theory can
be unambiguously reconstructed from two observable cosmological
functions of the redshift: the luminosity distance and the linear
density perturbation of dustlike matter. We also analyze the
constraints imposed on the theory by the knowledge of only the first
of these functions, as it will probably be available sooner with a
good accuracy.}

Theoretical attempts at quantizing gravity or at unifying it with
other interactions generically predict the existence of scalar
partners to the graviton. For instance, in superstring theory, a
dilaton is already present in the supermultiplet of the
10-dimensional graviton, and many other scalar fields, called the
``moduli'', appear when performing a Kaluza-Klein dimensional
reduction down to our usual spacetime. Gravity theories incorporating
such scalar fields have a great importance in cosmology, notably in
inflationary scenarios, but also for ``quintessence'' models, in
which the cosmological constant suggested by observations of type-Ia
supernovae appears in fact as the present value of a scalar-field
potential $U(\Phi)$.

Starobinsky\,\cite{St98} proved, in the case of a
minimally-coupled scalar field, that the precise shape of this
potential can be reconstructed if one knows either the present energy
density of dustlike matter $\Omega_{m,0}$ and the luminosity distance
$D_L(z)$ as a function of the redshift $z$, or the Hubble constant
$H_0$ and the density perturbation of dustlike matter $\delta_m(z)$
as a function of $z$. However, the models inspired by
extra-dimensional theories generically involve a non-minimal coupling
between the scalar field and the curvature. Several parametrizations
of the action may be chosen,\cite{beps00,ep00} but it can always be
written in the following ``Brans-Dicke'' form:
\begin{equation}
S = \int d^4 x \sqrt{-g}\left\{\Phi R - {\omega(\Phi)\over
\Phi}\left(\partial_\mu\Phi\right)^2-2 U(\Phi)\right\}
+ S_{\rm matter}\left[{\rm matter}; g_{\mu\nu}\right]\ ,
\end{equation}
where ``matter'' denotes all kinds of matter fields, including gauge
bosons (we consider here theories satisfying the ``weak equivalence
principle''). Since this action involves two unknown functions of the
scalar field, $\omega(\Phi)$ and $U(\Phi)$, we now need both
observable functions $D_L(z)$ and $\delta_m(z)$ to fully reconstruct
the theory. On the other hand, the present values of $\Omega_{m,0}$
and $H_0$ are not necessary.

The first step consists in obtaining the Hubble function $H(z)$ from
the knowledge of $D_L(z)$, like in general relativity since this is
purely kinematical: $1/H(z) = [D_L(z)/(1+z)]'$, where a prime denotes
a derivative with respect to $z$ (and where a corrective factor
involving $\Omega_{\kappa,0}\equiv -\kappa/(a_0H_0)^2$ enters if the
universe is spatially curved\,\cite{ep00}). The perturbation equations
in the limit of short wavelengths and the background
Friedmann-Robertson-Walker equations then give
\begin{eqnarray}
{\Phi\over \Phi_0} &\simeq& {3\over2}\left({H_0\over H}\right)^2
{(1+z)\,\Omega_{m,0}\,\delta_m\over \delta_m''+[H'/H -
(1+z)^{-1}]\delta_m'}
\times \left(1+{1\over 2\omega+3}\right)\ ,
\label{eq2}\\
{2U\over (1+z)^2H^2} &=&
\Phi'' + \left({H'\over H}- {4\over 1+z}\right)\Phi'
+\left[{6\over (1+z)^2} - {2\over 1+z}\,{H'\over H}
- 4 \left({H_0\over H}\right)^2\! \Omega_{\kappa,0}\right]\Phi
\nonumber\\
&& - 3\,(1+z)\left({H_0\over H}\right)^2\Phi_0\,\Omega_{m,0}\ ,
\label{eq3}\\
\omega &=& -{\Phi\over \Phi'^2}
\Biggl\{\Phi''
+ \left({H'\over H} + {2\over 1+z}\right)\Phi'
-2\left[{1\over 1+z}\,{H'\over H}
- \left({H_0\over H}\right)^2\! \Omega_{\kappa,0}\right]\Phi
\nonumber\\
&& + 3\,(1+z) \left({H_0\over H}\right)^2 \Phi_0\,\Omega_{m,0}
\Biggr\}\ ,
\label{eq4}
\end{eqnarray}
where the integration constant $\Phi_0$ may be set to 1 without loss
of generality. The three functions $\Phi(z)$, $U(z)$ and $\omega(z)$
--and thereby $U(\Phi)$ and $\omega(\Phi)$-- can thus be reconstructed
unambiguously if $D_L(z)$ and $\delta_m(z)$ are experimentally
determined precisely enough. Since solar-system experiments
constrain the present value of $\omega$ to be larger than 2500,
and that it can be proven\,\cite{beps00} to be larger that $\sim 5$
even for redshifts $z\sim1$, one can further simplify Eq.~(\ref{eq2})
by suppressing its last factor inside parentheses. Note that in that
case, the three equations become algebraic.

As $D_L(z)$ will be observed sooner with a good accuracy, we analyzed
the constraints imposed on scalar-tensor theories by this single
function.\cite{ep00} We assumed particular forms for either
$\omega(\Phi)$ or $U(\Phi)$, and reconstructed the other one thanks
to Eqs.~(\ref{eq3})-(\ref{eq4}) above. Our main conclusion is that
the knowledge of $D_L(z)$ over a wide redshift interval, say up to
$z \sim 2$, is sufficient to distinguish these theories from general
relativity plus a cosmological constant, even if there are large
(tens of percents) experimental errors. The reason for this
strong result is that we took into account not only solar-system (and
binary-pulsar) constraints, but also the following important
{\it theoretical\/} constraints: $\bullet$~The graviton should carry
positive energy ($\Rightarrow \Phi > 0$). $\bullet$~The scalar
field should carry positive energy ($\Rightarrow 2\omega(\Phi)+3 >
0$). $\bullet$~The potential $U(\Phi)/\Phi^2$ should be bounded by
below and have a reasonable shape, and to get a stable theory, the
scalar mass squared should be positive
($\Rightarrow \{\Phi[2\omega(\Phi)+3]^{-1/2}[U(\Phi)/\Phi^2]'\}'\geq
0$, where a prime denotes here $d/d\Phi$).


\section*{References}
\begin{thebibliography}{99}
\bibitem{St98}A.A.~Starobinsky, {\em JETP Lett.} {\bf 68}, 757 (1998);
{\em Gravitation Cosmol.} (Suppl.) {\bf 4}, 88 (1998).
\bibitem{beps00}B.~Boisseau, G.~Esposito-Far\`ese, D.~Polarski,
and A.A.~Starobinsky, {\em Phys. Rev. Lett.} {\bf 85}, 2236 (2000).
\bibitem{ep00}G.~Esposito-Far\`ese and D.~Polarski,
gr-qc/0009034, {\em Phys. Rev.} D (in press).
\end{thebibliography}
\end{document}